\documentclass{article}%
\usepackage{amsfonts}
\usepackage{amsmath}
\usepackage{amssymb}
\usepackage{graphicx}%
\providecommand{\U}[1]{\protect\rule{.1in}{.1in}}

\begin{document}

\title{ \textbf{Energy spectra of Hartmann and ring-shaped oscillator potentials
using the quantum Hamilton-Jacobi formalism}}
\author{A. GHARBI\thanks{Electronic address: \texttt{hakimgharbi@yahoo.fr}} , A.
BOUDA\thanks{Electronic address: \texttt{bouda\_a@yahoo.fr}}\\Laboratoire de Physique Th\'eorique, Universit\'e de B\'eja\"\i a,\\Campus Targa Ouzemour, 06000 B\'{e}ja\"{\i}a, Algeria\\}
\date{\today}
\maketitle

\begin{abstract}
\noindent In the present work, we apply the exact quantization condition,
introduced within the framework of Padgett and Leacock's quantum
Hamilton-Jacobi formalism, to angular and radial quantum action variables in
the context of the Hartmann and the ring-shaped oscillator potentials which
are separable and non central. The energy spectra of the two systems are
exactly obtained.

\end{abstract}

\vskip\baselineskip

\noindent\textbf{PACS:} 03.65.Ca

\noindent\textbf{Key words:} quantum Hamilton-Jacobi formalism, Hartmann
potential, ring-shaped oscillator potential, energy spectra.

\section{Introduction}

Since the advent of quantum mechanics, several methods have been developed in
order to find the exact energy spectra of bound states in stationary quantum
systems . The knowledge of these spectra is necessary for several applications
in many fields of physics and theoretical chemistry. Among these methods we
may mention the factorization method \cite{r1}, supersymmetric quantum
mechanics \cite{r2}, the integral equation method \cite{r3}, the path integral
formalism \cite{r4}, the momentum space method \cite{r5}, Ma and Xu's method
\cite{r6}, the group-theoretical method \cite{r6a} ...etc.

Leacock and Padgett proposed in 1983 \cite{r7,r8} a quantum version of the
Hamilton-Jacobi formalism where the energy spectrum of a stationary quantum
systems is obtained using an exact quantization condition. They postulated
that the Quantum Hamilton-Jacobi equation (QHJE) for a 1D stationary system is
of the form%
\begin{equation}
\frac{\hslash}{i}\frac{\partial^{2}W\left(  x,E\right)  }{\partial x^{2}%
}+\left(  \frac{\partial W\left(  x,E\right)  }{\partial x}\right)  ^{2}%
=2\mu\left[  E-V(x)\right]  \label{a1}%
\end{equation}
where $W\left(  x,E\right)  $ is the quantum Hamilton's characteristic
function, $V(x)$ is the potential, $E$ and $\mu$ are respectively the energy
and the mass of the particle. By defining the quantum momentum function (QMF)
as%
\begin{equation}
p\left(  x,E\right)  =\frac{\partial W\left(  x,E\right)  }{\partial x}
\label{a2}%
\end{equation}
equation $\left(  \ref{a1}\right)  $ becomes%
\begin{equation}
\frac{\hslash}{i}\frac{\partial p\left(  x,E\right)  }{\partial x}+p\left(
x,E\right)  ^{2}=2\mu\left[  E-V(x)\right]  =p_{c}\left(  x,E\right)  ^{2}
\label{a3}%
\end{equation}
where $p_{c}$ is the classical momentum function. Consequently we obtain the
following boundary \ condition on the QMF%
\begin{equation}%
\displaystyle
p\left(  x,E\right)  \underset{\hslash\rightarrow0}{\longrightarrow}%
p_{c}\left(  x,E\right)  \label{a4}%
\end{equation}
The previous condition can be considered as a correspondence principle
\cite{r7,r8}. By analogy to the classical action variable \cite{r9}, the
quantum action variable is defined in the complex $x$-plane by the following
contour integral
\begin{equation}
J=(1/2\pi)%
{\displaystyle\oint\limits_{c}}
dx\text{ }p\left(  x,E\right)  \label{a5}%
\end{equation}
where $c$ is a counterclockwise contour that encloses the two physical turning
points of $p_{c}\left(  x,E\right)  $. Padgett and Leacock \cite{r7,r8} have
shown that the quantum action variable $J$ can be used to obtain the energy
spectra through the following exact quantization condition%
\begin{equation}
J=n\hbar\text{ \ \ }n=0,1,2,\ldots\label{a6}%
\end{equation}
without solving equation $\left(  \ref{a3}\right)  $.

Kapoor \textit{et al}. derived energy eigenvalues for a class of
one-dimensional potentials \cite{r9a,r9b} and showed that in addition to
obtaining eigenenergies, Leacock and Padgett's formulation can also yield the
eigenfunctions of the energy \cite{r9c}. Subsequently a relativistic extension
of this approach was proposed by Kim and Choi \cite{r10,r11} where energy
spectra of some relativistic systems have been obtained. The Quantum
Hamilton-Jacobi formalism (QHJF) was also successfully applied to PT symmetric
hamiltonians and non hermitian exponential-type potentials \cite{r12,r13},
supersymmetric potentials \cite{r14},\ the position-dependent mass model
\cite{r15}, and two dimensional central potentials \cite{r15a} and two
dimensional singular oscillator \cite{r15b}.

Recently, the study of non central and separable potentials has sparked a
renewed interest of the research community, given their wide application in
quantum chemistry and nuclear physics. We would like to investigate the
application of Padgett's and Leacock's method to this type of potentials,
motivated by its plausible general applicability to all separable potentials.

In the present work, we apply the Padgett's and Leacock's approach to obtain
the energy spectra of two separable and non central potentials. The first one
is the Hartmann potential \cite{r16} which is obtained by adding to the
three-dimensional Coulombic term a ring-shaped inverse square one. It was
proposed in 1972 in quantum chemistry to describe ring-shaped molecules like
benzene. The second one is the ring-shaped oscillator which was introduced in
1988 by Quesne \cite{r17} through replacing the coulomb part of the Hartmann
potential by a harmonic oscillator term.

The manuscript is\ organized as follows. In section 2, the QHJE of a separable
and non-central potential is exposed in spherical polar coordinates. In
sections 3 and 4, the quantization condition is applied for angular and radial
quantum action variables in the context of the Hartmann and ring shaped
potentials respectively, and then the energy spectra of the two systems are
exactly obtained. In section 5, concluding remarks are given.

\section{Quantum Hamilton-Jacobi equation for a separable non central
potential}

The 3D Schr\"{o}dinger equation for a stationary system is given by:%
\begin{equation}
-\frac{\hbar^{2}}{2\mu}\Delta\psi\left(  \vec{r}\right)  +V\left(  \vec
{r}\right)  \psi\left(  \vec{r}\right)  =E\psi\left(  \vec{r}\right)
\label{c1}%
\end{equation}
where $\Delta$ is the Laplacian operator. The most general expression of a
potential for which equation $\left(  \ref{c1}\right)  $ is separable in
spherical polar coordinates is
\begin{equation}
V\left(  r,\theta,\phi\right)  =V_{1}(r)+\frac{V_{2}(\theta)}{r^{2}}%
+\frac{V_{3}(\phi)}{r^{2}\sin^{2}\theta} \label{c2}%
\end{equation}
where $V_{1}(r)$ , $V_{2}(\theta)$ , $V_{3}(\phi)$ are respectively arbitrary
functions of $r,\theta$ and $\phi$ \cite{r18,r18a,r19}. Writing the wave
function as
\begin{equation}
\psi=\frac{R\left(  r\right)  }{r}\frac{H\left(  \theta\right)  }{\left(
\sin\theta\right)  ^{\frac{1}{2}}}K\left(  \phi\right)  , \label{c3}%
\end{equation}
equation $\left(  \ref{c1}\right)  $ is separated (in units of $2\mu=1$) into
\begin{align}
\frac{d^{2}R\left(  r\right)  }{dr^{2}}+\frac{1}{\hbar^{2}}\left(
E-V_{1}(r)-\frac{\hbar^{2}\left(  l^{2}-\frac{1}{4}\right)  }{r^{2}}\right)
R\left(  r\right)   &  =0\label{c4}\\
\frac{d^{2}H\left(  \theta\right)  }{d\theta^{2}}-\frac{1}{\hbar^{2}}\left(
V_{2}(\theta)-\hbar^{2}\left(  l^{2}-\frac{m^{2}-\frac{1}{4}}{\sin^{2}\theta
}\right)  \right)  H\left(  \theta\right)   &  =0\label{c5}\\
\frac{d^{2}K\left(  \phi\right)  }{d\phi^{2}}+\left(  m^{2}-\frac{V_{3}(\phi
)}{\hbar^{2}}\right)  K\left(  \phi\right)   &  =0 \label{c6}%
\end{align}
where $l^{2}$ and $m^{2}$ are separation constants. We note that the three
previous equations themselves are Schr\"{o}dinger-like.

The generalization of the QHJE to three dimensions is given by (in units of
$2\mu=1$)
\begin{equation}
\frac{\hbar}{i}\Delta W\left(  \vec{r}\right)  +\vec{\nabla}W\left(  \vec
{r}\right)  \cdot\vec{\nabla}W\left(  \vec{r}\right)  =E-V\left(  \vec
{r}\right)  \label{c7}%
\end{equation}
where $\vec{\nabla}$ is the gradient operator, $W\left(  \vec{r}\right)  $ is
the Hamilton's characteristic function which is related to the solution of
Schr\"{o}dinger equation $\psi\left(  \vec{r}\right)  $ by \cite{r7,r8}
\begin{equation}
\psi\left(  \vec{r}\right)  =\exp\left(  iW\left(  \vec{r}\right)
/\hbar\right)  \label{c71}%
\end{equation}

From the definition of the QMF in three dimensions which is given by
\begin{equation}
\vec{p}=\vec{\nabla}W\left(  \vec{r}\right)  \label{c72}%
\end{equation}
one can find that the relation between the QMF and the wave function is of the
following form
\begin{equation}
\vec{p}=\frac{\hbar}{i}\frac{\vec{\nabla}\psi}{\psi} \label{c73}%
\end{equation}

By writing the characteristic function in spherical polar coordinates as
\begin{equation}
W\left(  \vec{r}\right)  =W_{r}\left(  r\right)  +W_{\theta}\left(
\theta\right)  +W_{\phi}\left(  \phi\right)  , \label{c7a}%
\end{equation}
we obtain from $\left(  \ref{c7}\right)  $ and $\left(  \ref{c2}\right)  $ the
following equations
\begin{align}
\frac{\hbar}{i}\left(  \frac{\partial^{2}W_{r}}{\partial r^{2}}+\frac{2}%
{r}\frac{\partial W_{r}}{\partial r}\right)  +\left(  \frac{\partial W_{r}%
}{\partial r}\right)  ^{2}  &  =E-V_{1}(r)-\frac{a}{r^{2}},\label{c8}\\
\frac{\hbar}{i}\left(  \frac{\partial^{2}W_{\theta}}{\partial\theta^{2}}%
+\frac{1}{\tan\theta}\frac{\partial W_{\theta}}{\partial\theta}\right)
+\left(  \frac{\partial W_{\theta}}{\partial\theta}\right)  ^{2}  &
=a-\frac{b}{\sin^{2}\theta}-V_{2}(\theta),\label{c9}\\
\frac{\hbar}{i}\frac{\partial^{2}W_{\phi}}{\partial\phi^{2}}+\left(
\frac{\partial W_{\phi}}{\partial\phi}\right)  ^{2}  &  =b-V_{3}(\phi),
\label{c10}%
\end{align}
where $a$ and $b$ are separation constants. Using the definitions
\cite{r7,r8}
\begin{equation}
p_{r}=\frac{\partial W_{r}}{\partial r},\text{ }p_{\theta}=\frac{\partial
W_{\theta}}{\partial\theta},\text{ }p_{\phi}=\frac{\partial W_{\phi}}%
{\partial\phi} \label{c11}%
\end{equation}
equations $\left(  \ref{c8}\right)  $, $\left(  \ref{c9}\right)  $ and
$\left(  \ref{c10}\right)  $ can be written as
\begin{align}
\frac{\hbar}{i}\left(  \frac{\partial p_{r}}{\partial r}+\frac{2}{r}%
p_{r}\right)  +p_{r}^{2}  &  =E-V_{1}(r)-\frac{a}{r^{2}}\label{c12}\\
\frac{\hbar}{i}\left(  \frac{\partial p_{\theta}}{\partial\theta}%
+\frac{p_{\theta}}{\tan\theta}\right)  +p_{\theta}^{2}  &  =a-\frac{b}%
{\sin^{2}\theta}-V_{2}(\theta)\label{c13}\\
\frac{\hbar}{i}\frac{\partial p_{\phi}}{\partial\phi}+p_{\phi}^{2}  &
=b-V_{3}(\phi) \label{c14}%
\end{align}

Later on, we will use the above equations in the application of the exact
quantization condition $\left(  \ref{a6}\right)  $ for each degree of freedom
and that for the Hartmann and the ring shaped potentials.

\section{Hartmann potential}

Hartmann potential was introduced in quantum chemistry as a model for some
ring shaped molecules. Its expression is
\begin{equation}
V\left(  r,\theta\right)  =\frac{\alpha}{r}+\frac{\beta}{r^{2}\sin^{2}\theta}
\label{d1}%
\end{equation}
where $\alpha$ and $\beta$ are real constants \cite{r16,r18,r18a}. Hence,
according to $\left(  \ref{c2}\right)  $ we have:%
\begin{equation}
V_{1}\left(  r\right)  =\frac{\alpha}{r}\text{ , }V_{2}\left(  \theta\right)
=\frac{\beta}{\sin^{2}\theta}\text{ , }V_{3}\left(  \phi\right)  =0.
\label{d2}%
\end{equation}
Let us start with the $\phi$ variable. Equation $\left(  \ref{c14}\right)  $
becomes%
\begin{equation}
\frac{\hbar}{i}\frac{\partial p_{\phi}}{\partial\phi}+p_{\phi}^{2}=b
\label{d2a}%
\end{equation}
We can easily show that the solution of $\left(  \ref{c6}\right)  $ is%
\begin{equation}
K\left(  \phi\right)  =e^{im\phi}\text{ } \label{d3}%
\end{equation}
and from the periodic boundary condition%
\begin{equation}
K\left(  \phi\right)  =K\left(  \phi+2\pi\right)  , \label{d3a}%
\end{equation}
one can find that%
\begin{equation}
m=0,\pm1,\pm2,\cdots\label{d3b}%
\end{equation}
So, using $\left(  \ref{c71}\right)  $, $\left(  \ref{c7a}\right)  $, $\left(
\ref{c11}\right)  $ and $\left(  \ref{d2a}\right)  $ we obtain%
\begin{equation}
b=\hbar^{2}m^{2}\rightarrow\sqrt{b}=\hbar m \label{d4}%
\end{equation}

For the $\theta$ variable, equation $\left(  \ref{c13}\right)  $ takes the
form%
\begin{equation}
\frac{\hbar}{i}\left(  \frac{\partial p_{\theta}}{\partial\theta}%
+\frac{p_{\theta}}{\tan\theta}\right)  +p_{\theta}^{2}=a-\frac{b+\beta}%
{\sin^{2}\theta}\label{d5}%
\end{equation}
{}

The $\theta$ quantum action variable is given by
\begin{equation}
J_{\theta}=\frac{1}{2\pi}%
{\displaystyle\oint\limits_{c_{\theta}}}
d\theta p_{\theta}, \label{d6}%
\end{equation}
where $c_{\theta}$ \cite{r7,r8} is a counterclockwise contour which encloses
the two turning points defined by the vanishing of the right-hand side of
$\left(  \ref{d5}\right)  $. After the change of variable $y=-\cot\theta$,
relations $\left(  \ref{d5}\right)  $ and $\left(  \ref{d6}\right)  $ become
respectively
\begin{equation}
\frac{\hbar}{i}\left(  \left(  y^{2}+1\right)  \frac{\partial p_{y}}{\partial
y}-yp_{y}\right)  +p_{y}^{2}=a-\beta-b-y^{2}\left(  \beta+b\right)  ,
\label{d7}%
\end{equation}
and
\begin{equation}
J_{\theta}=\frac{1}{2\pi}%
{\displaystyle\oint\limits_{c_{y}}}
\frac{p_{y}dy}{\left(  y+i\right)  \left(  y-i\right)  }, \label{d8}%
\end{equation}
where the integration is around the counterclockwise contour $c_{y}$ enclosing
the two turning points which are solutions of $a-\beta-b-y^{2}\left(
\beta+b\right)  =0,$ and the section of $\operatorname{Re}y$ axis between
them. The previous integral can be evaluated by distorting the contour $c_{y}$
to enclose the first order poles of the integrand at $y=\pm i$ and $y=\infty$
\cite{r7,r8}. Then we write%
\begin{equation}
J_{\theta}=J_{i}+J_{-i}+J_{\infty} \label{d9}%
\end{equation}

To evaluate $J_{\infty}$ at the pole $y=\infty$, we define a new variable
$s=1/y$, and we deduce%
\begin{equation}
J_{\infty}=\frac{1}{2\pi}%
{\displaystyle\oint\limits_{c_{s}}}
\frac{p_{s}ds}{\left(  s-i\right)  \left(  s+i\right)  }, \label{d17}%
\end{equation}
where $c_{s}$ is a counterclockwise contour that encloses only the pole at
$s=0$. Equation $\left(  \ref{d7}\right)  $ becomes%
\begin{equation}
-\frac{\hbar}{i}\left(  \left(  1+s^{2}\right)  \frac{\partial p_{s}}{\partial
s}+\frac{p_{s}}{s}\right)  +p_{s}^{2}=a-\beta-b-\frac{\left(  \beta+b\right)
}{s^{2}} \label{d18}%
\end{equation}
We write $p_{s}$ near $s=0$ as \cite{r7,r8}
\begin{equation}
p_{s}=\frac{b_{1}}{s}+a_{0}+a_{1}s+... \label{d18a}%
\end{equation}
Substituting this expression into equation $\left(  \ref{d18}\right)  $, we
deduce that
\begin{equation}
b_{1}=\pm i\sqrt{\beta+b}. \label{d19}%
\end{equation}
We note that it is only the term proportional to $b_{1}$ in $\left(
\ref{d18a}\right)  $ which has a non vanishing contribution in $\left(
\ref{d17}\right)  $. To remove the ambiguity in the sign of $\ b_{1}$, we
apply the boundary condition $\left(  \ref{a4}\right)  $. Indeed, according to
\cite{r7,r8,r9a}, the classical momentum function $p_{s}^{c}$\ is defined as
the branch of the square root which is positive just below the branch cut
joining the two classical turning points given by the solution of the equation
$p_{s}^{c}=0$. Therefore, from $\left(  \ref{d18}\right)  $\ , the classical
momentum function satisfies $p_{s}^{c}=+i\left(  \beta+b\right)  ^{1/2}/s$
near$\ s=0$ ($y=\infty$). It follows from $\left(  \ref{d18a}\right)  $ that
$b_{1}=+i\left(  \beta+b\right)  ^{1/2}$ . Applying the residue theorem to
$\left(  \ref{d17}\right)  $, we find%
\begin{equation}
J_{\infty}=\frac{1}{2\pi}2\pi ib_{1}=-\sqrt{\beta+b} \label{d20}%
\end{equation}

To evaluate $J_{i}$ we make a change of variable $y=s+i$, and we obtain%
\begin{equation}
J_{i}=\frac{1}{2\pi}%
{\displaystyle\oint\limits_{c_{s}}}
\frac{p_{s}ds}{\left(  s+2i\right)  s}, \label{d10}%
\end{equation}
where $c_{s}$ is a clockwise contour that encloses only the pole at $s=0$.
Equation $\left(  \ref{d7}\right)  $becomes
\begin{equation}
\frac{\hbar}{i}\left(  \allowbreak s^{2}+2is\right)  \frac{\partial p_{s}%
}{\partial s}-\frac{\hbar}{i}\left(  s+i\right)  p_{s}+p_{s}^{2}=a-\left(
b+\beta\right)  s^{2}-\allowbreak2i\left(  \beta+b\right)  s \label{d11}%
\end{equation}
We expand $p_{s}$ near $s=0$ as
\begin{equation}
p_{s}=\frac{b_{1}}{s}+a_{0}+a_{1}s+... \label{d12}%
\end{equation}
and note that the only contributing coefficient in $\left(  \ref{d12}\right)
$ to the contour integral $\left(  \ref{d10}\right)  $ is $a_{0}$.
Substituting $\left(  \ref{d12}\right)  $ in $\left(  \ref{d11}\right)  $ and
matching terms of same power of $s$, we obtain%
\begin{equation}
a_{0}=\frac{\hbar}{2}\pm\sqrt{a+\left(  \frac{\hbar}{2}\right)  ^{2}}.
\label{d13}%
\end{equation}
According to refs \cite{r7,r8}, the ambiguity in the sign of $a_{0}$ can be
removed by using the boundary condition $\left(  \ref{a4}\right)  $. Then we
find%
\begin{equation}
a_{0}=\frac{\hbar}{2}-\sqrt{a+\left(  \frac{\hbar}{2}\right)  ^{2}}.
\label{d14}%
\end{equation}
So, using the residue theorem we find%
\begin{equation}
J_{i}=-\frac{a_{0}}{2}=-\frac{\hbar}{4}+\frac{1}{2}\sqrt{a+\left(  \frac
{\hbar}{2}\right)  ^{2}}. \label{d15}%
\end{equation}
The contribution $J_{-i}$ at the pole $y=-i$ is evaluated in a similar way as
$J_{i}$ and we obtain%
\begin{equation}
J_{-i}=\frac{1}{2}\sqrt{a+\left(  \frac{\hbar}{2}\right)  ^{2}}-\frac{\hbar
}{4}. \label{d16}%
\end{equation}
. Then from $\left(  \ref{d20}\right)  ,\left(  \ref{d15}\right)  $ and
$\left(  \ref{d16}\right)  $, we find
\begin{equation}
J_{\theta}=\sqrt{\left(  \frac{\hbar}{2}\right)  ^{2}+a}-\sqrt{b+\beta}%
-\frac{\hbar}{2} \label{d21}%
\end{equation}

Using the quantization condition for $\theta$ variable, we obtain%
\begin{equation}
J_{\theta}=\sqrt{\left(  \frac{\hbar}{2}\right)  ^{2}+a}-\sqrt{b+\beta}%
-\frac{\hbar}{2}=\hbar n_{\theta}\text{ \ \ ,\ }n_{\theta}=0,1,2,\ldots
\label{d22}%
\end{equation}

For $r$ variable, equation $\left(  \ref{c12}\right)  $\ takes the form%
\begin{equation}
\frac{\hbar}{i}\left(  \frac{\partial p_{r}}{\partial r}+\frac{2}{r}%
p_{r}\right)  +p_{r}^{2}=E-\frac{\alpha}{r}-\frac{a}{r^{2}} \label{d23}%
\end{equation}
The $r$ quantum action variable is given by \cite{r7,r8}
\begin{equation}
J_{r}=\frac{1}{2\pi}%
{\displaystyle\oint\limits_{c_{r}}}
drp_{r}\text{ ,} \label{d24}%
\end{equation}
where $c_{r}\ $is a counterclockwise contour that encloses the two turning
points which are solutions of%
\[
E-\frac{\alpha}{r}-\frac{a}{r^{2}}=0
\]
and the section of $\operatorname{Re}r$ axis between them. The integral
$\left(  \ref{d24}\right)  $ may be evaluated by distorting $c_{r}$ to enclose
the poles of the integrand at $r=0$ and $r=\infty$. We call $J_{0}$ and
$J_{\infty}$ the contributions to $J_{r}$ at $r=0$ and $r=\infty$
respectively. Then, we write%
\begin{equation}
J_{r}=J_{0}+J_{\infty} \label{d25}%
\end{equation}

To calculate $J_{0}$ we let $p_{r}=b_{1}/r+a_{0}+a_{1}r+...$ near $r=0$ and
replace this form of $p_{r}$ in $\left(  \ref{d23}\right)  $. Comparing the
coefficients of the $1/r^{2}$ terms gives%
\begin{equation}
b_{1}=i\frac{\hbar}{2}\pm i\sqrt{\left(  \frac{\hbar}{2}\right)  ^{2}+a}
\label{d26}%
\end{equation}
As the classical momentum function near $r=0$ is given by $p_{r}^{c}%
=-i\sqrt{a}/r$ , according to refs \cite{r7,r8,r15a}, the boundary condition
$\left(  \ref{a4}\right)  $ indicates that the lower sign in $b_{1}$ is the
correct one. Hence, we find%
\begin{equation}
b_{1}=i\frac{\hbar}{2}-i\sqrt{\left(  \frac{\hbar}{2}\right)  ^{2}+a}
\label{d27}%
\end{equation}

Applying the residue theorem to evaluate $J_{0}$, where the distorted contour
around $r=0$ is in the clockwise direction, gives%
\begin{equation}
J_{0}=\frac{1}{2\pi}\left(  -2\pi i\right)  b_{1}=\frac{\hbar}{2}%
-\sqrt{\left(  \frac{\hbar}{2}\right)  ^{2}+a}. \label{d28}%
\end{equation}
To evaluate $J_{\infty}$ we make the change of variable $s=1/r$, and find
\begin{equation}
J_{\infty}=\frac{1}{2\pi}%
{\displaystyle\oint\limits_{c_{s}}}
\frac{p_{s}}{s^{2}}ds\text{ ,} \label{d29}%
\end{equation}
where $c_{s}$ is a counterclockwise contour enclosing the pole at $s=0$.
Equation $\left(  \ref{d23}\right)  $ becomes%
\begin{equation}
\frac{1}{i}\left(  -s^{2}\frac{\partial p}{\partial s}+2sp_{s}\right)
+p_{s}^{2}=E-\alpha s-as^{2}. \label{d30}%
\end{equation}
We expand $p_{s}$ near $s=0$ as $p_{s}=b_{1}/s+a_{0}+a_{1}s+...$ and
substitute it in the last equation. Collecting the coefficients of the zeroth
and the first power of $s$ terms on each side of $\left(  \ref{d30}\right)  $
we find that $a_{0}^{2}=E$ and $2ia_{0}-2a_{0}a_{1}=\alpha$. Using the
boundary condition $\left(  \ref{a4}\right)  $, we find $a_{1}=i\hbar
-i\alpha/2\sqrt{-E}$ \cite{r7,r8,r15a}. Applying the residue theorem to
evaluate $\left(  \ref{d29}\right)  $\ we obtain%
\begin{equation}
J_{\infty}=\frac{1}{2\pi}\left(  2\pi i\right)  a_{1}=\frac{\alpha}{2\sqrt
{-E}}-\hbar\label{d31}%
\end{equation}
From $\left(  \ref{d28}\right)  $ and $\left(  \ref{d31}\right)  $ and using
the quantization condition for $r$ variable, we can write%
\begin{equation}
J_{r}=\frac{\alpha}{2\sqrt{-E}}-\frac{\hbar}{2}-\sqrt{\left(  \frac{\hbar}%
{2}\right)  ^{2}+a}=\hbar n_{r}\ \ ,n_{r}=0,1,2,\ldots\label{d32a}%
\end{equation}
From $\left(  \ref{d4}\right)  $, $\left(  \ref{d22}\right)  $ and $\left(
\ref{d32a}\right)  $ we obtain
\begin{equation}
E_{n_{r},n_{\theta}}=-\frac{1}{4\hbar^{2}}\frac{\alpha^{2}}{\left(
n_{r}+n_{\theta}+\sqrt{\frac{\beta}{\hbar^{2}}+m^{2}}+1\right)  ^{2}}
\label{d33a}%
\end{equation}
where $n_{r},n_{\theta}=0,1,2,\dots$\ and\ $m=0,\pm1,\pm2,\dots$. Our result
$\left(  \ref{d33a}\right)  $\ is identical to the one obtained in refs
\cite{r18,r18a} for energy levels of the Hartmann potential.

Now we will point out how to obtain the energy eigenfunctions. For the $\phi$
variable, we have already established in $\left(  \ref{d3}\right)  $ and
$\left(  \ref{d3b}\right)  $ that
\begin{equation}
K\left(  \phi\right)  =e^{im\phi},\text{ \ }m=0,\pm1,\pm2,\cdots\label{d34}%
\end{equation}
For the $\theta$ variable, according to relations $\left(  \ref{c3}\right)  $,
$\left(  \ref{c71}\right)  $ and $\left(  \ref{c7a}\right)  $, we can write%
\begin{equation}
H\left(  \theta\right)  =\left(  \sin\theta\right)  ^{\frac{1}{2}}\exp
i\left(  \frac{W_{\theta}\left(  \theta\right)  }{\hbar}\right)  \label{d35}%
\end{equation}
Substituting the above relation in $\left(  \ref{c5}\right)  $, and comparing
with $\left(  \ref{c9}\right)  $, we find that
\begin{equation}
l^{2}=\frac{1}{\hbar^{2}}\left(  a+\left(  \frac{\hbar}{2}\right)
^{2}\right)  \label{d36}%
\end{equation}
and from $\left(  \ref{d4}\right)  $ and $\left(  \ref{d22}\right)  $ we
obtain%
\begin{equation}
l^{2}=\left(  n_{\theta}+\sqrt{m^{2}+\frac{\beta}{\hbar^{2}}}+\frac{1}%
{2}\right)  ^{2}. \label{d37}%
\end{equation}

Replacing the above result in $\left(  \ref{c5}\right)  $ and after a suitable
change of variable, equation $\left(  \ref{c5}\right)  $ is transformed into a
standard hypergeometric equation and then we can write the solution $H\left(
\theta\right)  $ in terms of a hypergeometric function (for more details see
ref \cite{r18a}).

For $r$ variable, we substitute $\left(  \ref{d37}\right)  $ and $\left(
\ref{d33a}\right)  $ in the radial Schr\"{o}dinger equation given by $\left(
\ref{c4}\right)  $, and through an appropriate change of variable, we obtain a
differential equation which is satisfied by Laguerre polynomials $L_{p}%
^{k}\left(  z\right)  $. So, the solution $R\left(  r\right)  $ can be written
in terms of Laguerre polynomials (see ref \cite{r18a}).

\section{Ring-shaped potential}

The ring shaped potential was proposed for the first time by Quesne in 1988
\cite{r17,r18a,r19}. Its expression in spherical coordinates is given by%
\begin{equation}
V\left(  r,\theta\right)  =\alpha r^{2}+\frac{\beta}{r^{2}\sin^{2}\theta}
\label{e1}%
\end{equation}
where $\alpha$ and $\beta$ are real constants. Comparing to $\left(
\ref{c2}\right)  $ we have%
\begin{equation}
V_{1}\left(  r\right)  =\alpha r^{2}\text{ , }V_{2}\left(  \theta\right)
=\frac{\beta}{\sin^{2}\theta}\text{ , }V_{3}\left(  \phi\right)  =0 \label{e2}%
\end{equation}
Therefore equations $\left(  \ref{c8}\right)  $, $\left(  \ref{c9}\right)  $
and $\left(  \ref{c10}\right)  $ become%
\begin{align}
\frac{\hbar}{i}\left(  \frac{\partial p_{r}}{\partial r}+\frac{2}{r}%
p_{r}\right)  +p_{r}^{2}  &  =E-\alpha r^{2}-\frac{a}{r^{2}}\label{e3}\\
\frac{\hbar}{i}\left(  \frac{\partial p_{\theta}}{\partial\theta}%
+\frac{p_{\theta}}{\tan\theta}\right)  +p_{\theta}^{2}  &  =a-\frac{b+\beta
}{\sin^{2}\theta}\label{e4}\\
\frac{\hbar}{i}\frac{\partial p_{\phi}}{\partial\phi}+p_{\phi}^{2}  &  =b
\label{e5}%
\end{align}

We note that equations $\left(  \ref{e4}\right)  $ and $\left(  \ref{e5}%
\right)  $ are respectively identical to equations $\left(  \ref{d5}\right)  $
and $\left(  \ref{d2a}\right)  $. Thus, the obtained results for $\phi$ and
$\theta$ variables for the Hartmann potential remain valid for the ring-shaped
potential. Accordingly, we apply the quantization condition $\left(
\ref{a6}\right)  $ for the ring-shaped potential only for the $r$ variable.

The $r$ quantum action variable is given by \cite{r7,r8}%
\begin{equation}
J_{r}=\frac{1}{2\pi}%
{\displaystyle\oint\limits_{c_{r}}}
drp_{r} \label{e6}%
\end{equation}
where $c_{r}$ is a counterclockwise contour enclosing the two physical turning
points which are solutions of%
\begin{equation}
E-\alpha r^{2}-\frac{a}{r^{2}}=0 \label{7e}%
\end{equation}
and the section of $\operatorname{Re}r$ axis between them. We note that about
the four solutions of $\left(  \ref{7e}\right)  $
\begin{align*}
r_{1}  &  =\left(  \frac{1}{2\alpha}\left(  E+\sqrt{-4a\alpha+E^{2}}\right)
\right)  ^{1/2},r_{3}=-r_{1}\\
r_{2}  &  =\left(  \frac{1}{2\alpha}\left(  E-\sqrt{-4a\alpha+E^{2}}\right)
\right)  ^{1/2},r_{4}=-r_{2}%
\end{align*}
we retain only the two positive ones, $r_{1}$ and $r_{2}$, the two others are
unphysical \cite{r7,r8,r15a}.

To calculate $J_{r}$, the contour $c_{r}$ is distorted to enclose the poles at
$r=0$ and $r=\infty$ as well as the additional poles of $p_{r}$ located on the
negative $\operatorname{Re}r$ axis between the unphysical turning points
$r_{3}$ and $r_{4}$. Then we write $J_{r}=J_{0}+J_{\infty}+J_{-}$ where
$J_{-}$ is the contribution to $J_{r}$ from the poles between $r_{3}$ and
$r_{4}$. Since the effective potential $V_{eff}=\alpha r^{2}+a/r^{2}$ is
symmetric with respect to $r=0$, we can associate to each pole of $p_{r}$
between $r_{1}$ and $r_{2}$ a symmetrical pole relative to $r=0$ with residue
$-i$ located between $r_{3}$ and $r_{4}$. In other words, $p_{r}$ is
symmetrical with respect to $r=0$ in number and location of poles with residue
$-i$. Because the distorted contour enclosing poles on the negative
$\operatorname{Re}r$ axis is clockwise,\ we have $J_{r}=-J_{-}$. Consequently
we obtain $J_{r}=\left(  J_{0}+J_{\infty}\right)  /2$ \cite{r7,r8,r15a}.
$J_{0}$ and $J_{\infty}$ are evaluated by following the same method used above
in the case of radial quantum action variable of Hartmann potential. We obtain
$J_{0}=\frac{1}{2}\hbar-\sqrt{\left(  \frac{\hbar}{2}\right)  ^{2}+a}$ and
$J_{\infty}=-\frac{3}{2}\hbar+\frac{1}{2}\frac{E}{\sqrt{\alpha}}$. Using the
quantization condition for $r$ variable, we find
\begin{equation}
J_{r}=\frac{1}{2}\left(  \frac{1}{2}\frac{E}{\sqrt{\alpha}}-\hbar
-\sqrt{\left(  \frac{\hbar}{2}\right)  ^{2}+a}\right)  =n_{r}\hbar\text{,
}n_{r}=0,1,2,\ldots\label{8e}%
\end{equation}
From $\left(  \ref{d4}\right)  $, $\left(  \ref{d22}\right)  $ and $\left(
\ref{8e}\right)  $ we reproduce the exact energy spectrum for the ring shaped
oscillator
\begin{equation}
E_{n_{r},n_{\theta}}=2\hbar\sqrt{\alpha}\left(  2n_{r}+n_{\theta}+\frac{3}%
{2}+\sqrt{\frac{\beta}{\hbar^{2}}+m^{2}}\right)  , \label{9e}%
\end{equation}
where $n_{r},n_{\theta}=0,1,2,\dots$\ and\ $m=0,\pm1,\pm2,\dots$
\cite{r18a,r19}.

In order to obtain the energy eigenfunctions of the ring shaped oscillator, we
proceed in the same way as explained for the Hartmann potential (see ref
\cite{r18a}).

\section{Conclusion}

In this work, we have obtained the exact energy spectra of Hartmann and ring
shaped oscillator potentials in the framework of Padegtt and Leacock's
formalism. After separation of variables for the tridimensional QHJE, and
using the definition of the QMF for $r$, $\theta$ and $\phi$ variables, we
apply the exact quantization condition to radial and angular quantum action
variables. This yields the energy levels for the two studied systems.

An important feature of this method resides in its ease of use. Indeed, to
arrive at eigenenergies, it does not require an analytical solution of the QMF
equation but merely the knowledge of its singularity structure.

As a follow-up, we intend to apply this method to other separable and non
central potentials in the context of relativistic equations such as
Klein-Gordon or Dirac equations.

\end{document}